 \documentclass[aps, prl,amsmath,amssymb,twocolumn,floatfix,showpacs]{revtex4} 
 \usepackage[usenames,dvipsnames]{color} 
 \usepackage[pdftex]{graphicx} 
 \usepackage{epsfig} 
 \input epsf 
 
\graphicspath{ {figs/} }

 \newcommand {\dr}{{\mathrm d}\mathbf{r}}

 \newcommand {\rr}{\mathbf{r}}

\begin{document} 
 
 \title{{Quasicrystalline order and a crystal-liquid state in a soft-core fluid}}
 
 \author{A.J.~Archer$^\ast$, A.M.~Rucklidge$^\dag$ and E.~Knobloch$^\#$} 
 \affiliation{$^\ast$Department of Mathematical Sciences, Loughborough University, Loughborough LE11 3TU, UK\\ 
$^\dag$Department of Applied Mathematics, University of Leeds, Leeds LS2 9JT, UK\\ 
$^\#$Department of Physics, University of California at Berkeley, Berkeley, CA 94720, USA} 
 
 \begin{abstract} 
A two-dimensional system of soft particles interacting via a two-length-scale 
potential is studied. Density functional theory and Brownian dynamics simulations reveal a 
fluid phase and two crystalline phases with different lattice spacing. Of 
these the larger lattice spacing phase can form an exotic 
{periodic state with a fraction of highly mobile particles: a crystal liquid}. Near the transition
between this phase and the smaller lattice spacing 
phase, quasicrystalline structures may
be created by a competition between linear instability at one scale and nonlinear selection of the
other.
 \end{abstract} 
 \maketitle 
 \epsfclipon 
 
Crystals are ordered arrangements of atoms or molecules with rotation 
and translation symmetries. Quasicrystals (QCs), discovered in 
1982~\cite{Shechtman1984a}, lack the lattice symmetries of crystals and yet 
have discrete Fourier spectra. QCs have been found not only in metals 
but also in colloidal systems~\cite{DentonLowen98,Talapin2009},  
mesoporous silica~\cite{Xiao2012} and soft-matter systems~\cite{Dotera2011}.  
The latter can form micelles~\cite{Zeng2004,Fischer2011}, e.g., from 
dendrimers or block copolymers, comprising a hydrophobic polymer 
core surrounded by a corona of flexible hydrophilic polymer chains. 
Theoretical approaches to investigating the stability of metallic or micellar 
QCs often involve minimising an appropriate energy, but the principle 
underlying their stability is not known~\cite{KeGl07, HBS06}. 

Patterns with quasicrystalline structure, or quasipatterns, were discovered in
Faraday wave experiments in the 1990s; two mechanisms for stabilizing these 
were identified \cite{Muller1994}. The first, relevant to experiments in
Ref.~\cite{Christiansen1992}, involves one length scale and may lead to a 
stable quasipattern \cite{Malomed1989}. The second, involving coupling between
an unstable scale and weakly damped (or weakly excited) waves with a different
length scale, is relevant to the experiments in 
Refs.~\cite{Edwards1993,Kudrolli1998,Arbell2002,Ding2006}, and was explored 
in~\cite{Zhang1997,Lifshitz1997,Topaz2004,Skeldon2007,Rucklidge2009,Rucklidge2012}. 
This mechanism can also operate for soft-matter QCs
\cite{Lifshitz2007a,Engel2007a,Barkan2011,RGZ12}. Here, we observe a
dynamic mechanism for forming QCs involving two length scales that is
qualitatively different: the system first forms a small length scale crystal.
Only when this phase is almost fully formed (i.e., the
dynamics is far into the nonlinear regime) does the longer length scale start
to appear, leading to the formation of QCs. This process occurs in a region of the 
phase diagram where the linear growth of density fluctuations in a quenched
uniform fluid selects the shorter scale but nonlinear stability favors a
longer scale.

The effective coarse-grained interaction potentials between the centres of 
mass of polymers, dendrimers or other such macromolecules, are soft. By this 
we mean that they are finite for all separation distances $r$, because the 
centre of mass of such soft objects does not necessarily coincide with any 
individual monomer. The soft effective pair potential between such particles 
can be approximated as $V(r)=\epsilon e^{-(r/R)^n}$. Simple linear polymers in solution 
correspond to the case $n=2$ with the length $R$ of order the radius of 
gyration and the energy $\epsilon$ for a pair of polymers to fully 
overlap of order $2k_BT$, where $k_B$ is Boltzmann's constant and $T$ is the 
temperature 
\cite{Likos01,DaHa94, likos:prl:98, BLHM01, likos:harreis:02, Likos06, LBLM12}. 
Dendrimers, due to the nature of their chemical architecture, can have an
effective interaction with a higher value of $n$; such systems form so-called
cluster crystals \cite{LBLM12} and there has been much interest in soft
potential models for these systems \cite{LLWL00, ALE04, 
MGKNL07, MoLi07, LMGK07, TMAL09, NKL12}.
 
Here we consider a model two-dimensional system of soft particles that
interact via the potential
 \begin{equation} 
 V(r)=\epsilon e^{-(r/R)^8}+\epsilon Ae^{-(r/R_s)^8}. 
 \label{eq:pair_pot} 
 \end{equation} 
This potential is finite for all $r$ and has a shoulder when 
the parameter $A\neq0$, with two length scales. The radius of the 
core is $R$ and the radius of the shoulder is $R_s>R$; the energy 
for complete overlap is $(1+A)\epsilon$. Such a potential is a 
simple coarse-grained model for the effective interaction between 
dendrimers, star polymers or micelles formed, e.g., from block copolymers, 
which have a stiff hydrophobic core surrounded by a corona of flexible 
hydrophilic chains. A related, but piecewise constant potential is used in 
Ref.\ \cite{Barkan2011}. The limits (i) $A\to0$ or (ii) $A\to \infty$ 
{\em and} $\epsilon\to 0$ with $\epsilon A=$ constant both result in
systems with a single crystal phase. 
In the following we set the dimensionless interaction energy parameter 
$\beta\epsilon=1$, where $\beta=(k_BT)^{-1}$, and fix the ratio of the 
two length scales to be $R_s/R=1.855$ (see below). 
 
We use density functional theory (DFT) \cite{Evans79,Evans92,Lutsko10} to study 
this system. The grand free energy is 
\begin{eqnarray}\notag 
 \Omega[\rho(\rr)]&=&k_BT \int \dr \rho(\rr)[\ln\Lambda^2\rho(\rr)-1]\\ 
         &\,& {}+ \mathcal{F}_{ex}[\rho(\rr)] +\int \dr \,(\Phi(\rr)-\mu)\rho(\rr), 
 \label{eq:DFT} 
 \end{eqnarray} 
which is a functional of the one-body (number) density of the particles, $\rho(\rr)$, 
where $\rr=(x,y)$. The first term is the ideal-gas contribution to the free energy $\mathcal{F}_{id}$, 
$\Lambda$ is the thermal de Broglie wavelength, $\mu$ is the 
chemical potential, $\Phi(\rr)$ is any external potential that may be 
confining the system, and $\mathcal{F}_{ex}[\rho(\rr)]$ is the 
excess Helmholtz free energy from the interactions between the particles. The 
equilibrium density profile is that which minimizes $\Omega[\rho(\rr)]$; 
the corresponding minimum is the thermodynamic grand potential of the system. 
For a system in the bulk fluid state (i.e., where $\Phi(\rr)\equiv 0$), the 
minimising density is uniform, $\rho=\rho_0$. However, for other state points, 
when the system freezes to form a solid, $\Omega$ is minimized by a nonuniform 
$\rho(\rr)$, exhibiting sharp peaks. For the systems of soft-core 
particles considered here, one may approximate $\mathcal{F}_{ex}$ as~\cite{Likos01}: 
 \begin{equation} 
 \mathcal{F}_{ex}[\rho(\rr)]=\frac{1}{2}\int\dr\int\dr'\rho(\rr)V(|\rr-\rr'|)\rho(\rr'). 
 \label{eq:Fex} 
 \end{equation} 
 This functional generates the random phase approximation (RPA) for the pair 
direct correlation function $c^{(2)}(\rr,\rr')\equiv-\beta\frac{\delta^2 
\mathcal{F}_{ex}}{\delta \rho(\rr) \delta \rho(\rr')}=-\beta V(|\rr-\rr'|)$ 
\cite{Evans79,Evans92,Lutsko10}. If we assume that these are Brownian particles 
with dynamics 
 \begin{equation} 
 \dot{\rr}_i= -\Gamma\nabla_i U(\{\rr_i\},t) + \Gamma{\bf X}_i(t), 
 \label{eq:EOM} 
 \end{equation} 
where the index $i=1,...,N$ labels particles, 
$U(\{\rr_i\},t)$ $=\sum_{i=1}^N\Phi(\rr_i)+\sum_{i\neq j}V(\rr_i-\rr_j)$ is the 
potential energy of the system and ${\bf X}_i(t)$ is a white noise term, 
we can investigate the dynamics of the system using Dynamic Density Functional Theory (DDFT)~\cite{MaTa99,MaTa00,ArEv04,ArRa04} in the form 
 \begin{equation} 
 \frac{\partial\rho(\rr,t)}{\partial t} = 
  \Gamma \nabla\cdot\left[\rho(\rr,t)\nabla\frac{\delta\Omega[\rho(\rr,t)]}{\delta\rho(\rr,t)}\right], 
 \label{eq:DDFT} 
 \end{equation} 
where $\rho(\rr,t)$ is now the time-dependent nonequilibrium one-body density 
profile and  $\Gamma\equiv\beta D$ is the mobility. Here $D$ is the diffusion 
coefficient. In deriving Eq.~\eqref{eq:DDFT} we have used the equilibrium 
free energy $\mathcal{F}=\mathcal{F}_{id}+\mathcal{F}_{ex}$ to approximate 
the unknown nonequilibrium free energy. 
 
 \begin{figure*} 
 \begin{minipage}[b]{0.20\hsize} 
 \includegraphics[width=\hsize]{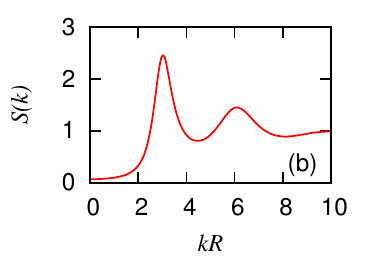} 
 \includegraphics[width=\hsize]{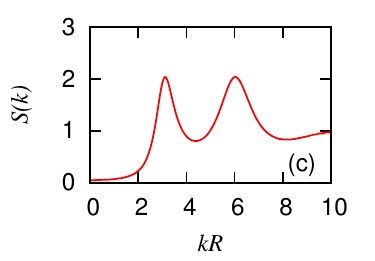} 
 \includegraphics[width=\hsize]{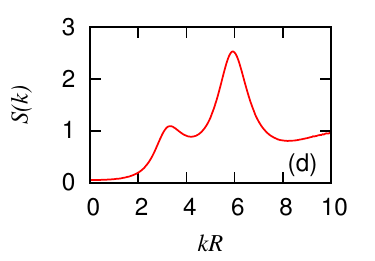}\\ \phantom{\small{xx}} 
 \end{minipage} 
 \includegraphics[width=0.55\hsize]{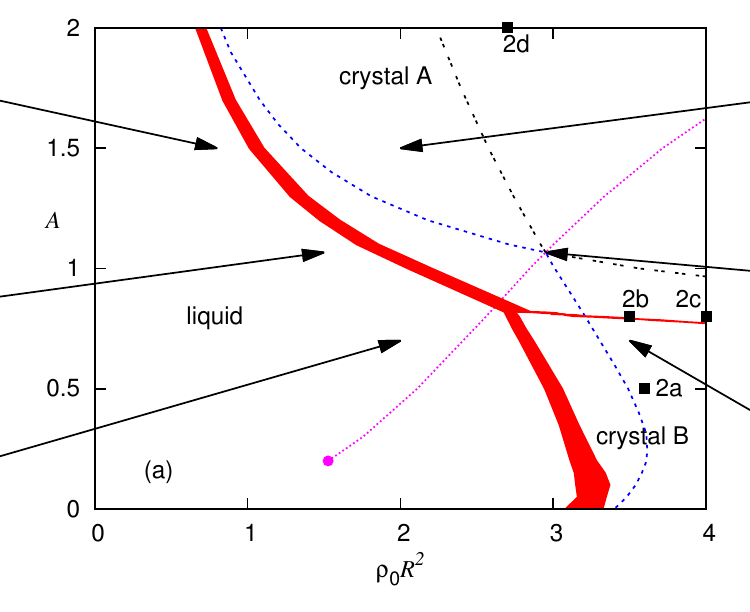} 
 \begin{minipage}[b]{0.20\hsize} 
 \includegraphics[width=\hsize]{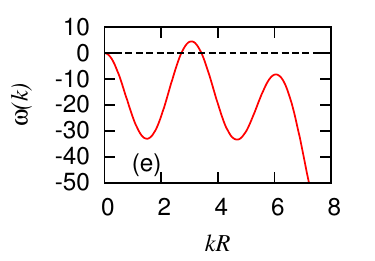} 
 \includegraphics[width=\hsize]{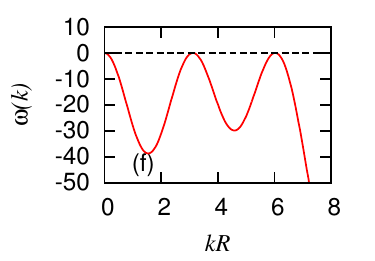} 
 \includegraphics[width=\hsize]{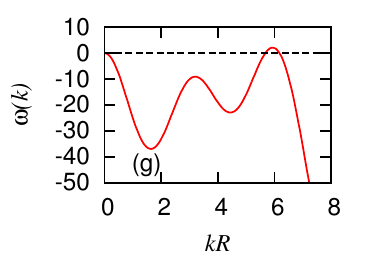}\\ \phantom{\small{xx}} 
 \end{minipage} 
\vspace{-2ex} 
 \caption{\label{fig:all} (color online) Phase diagram, static structure factor $S(k)$ 
and dispersion relation $\omega(k)$ for $\beta\epsilon=1$ and 
$R_s/R=1.855$. (a) The bulk system phase diagram in the 
$(\rho_0R^2,A)$ plane. The system exhibits a uniform fluid phase and two 
crystal phases: the larger lattice spacing crystal~A phase and the smaller 
lattice spacing crystal~B phase. The regions filled in red denote areas where 
there is two-phase coexistence between the different phases. The blue dashed 
line denotes the linear instability threshold for the liquid phase while the 
pink dotted line terminating in a circle is the locus where the two peaks in 
the dispersion relation \eqref{eq:disp_rel} have the same height. The circle 
denotes the point where the smaller $k$ peak disappears. (b)--(d) $S(k)$ for 
(b) $(\rho_0R^2,A)=(0.8,1.5)$, (c) $(1.5,1.067)$, (d) $(2,0.7)$. (e)--(g) $\omega(k)$ 
for (e) $(\rho_0R^2,A)=(2,1.5)$, (f) $(2.95,1.067)$, (g) $(3.5,0.7)$. 
The state points corresponding to the profiles in Fig.~\ref{fig:profiles} are marked with the symbol $\blacksquare$. 
\vspace{-1ex}} 
 \end{figure*} 
 
 \begin{figure*} 
 \hbox to \hsize{\hfil 
                 \hbox to 0.49\columnwidth{\hfil(a) $(3.6,0.5)$\hfil}%
                 \hfil 
                 \hbox to 0.49\columnwidth{\hfil(b) $(3.5,0.76)$\hfil}%
                 \hfil 
                 \hbox to 0.49\columnwidth{\hfil(c) $(4.0,0.8)$\hfil}%
                 \hfil 
                 \hbox to 0.49\columnwidth{\hfil(d) $(2.7,2)$\hfil}%
                 \hfil} 
 \includegraphics[width=0.49\columnwidth]{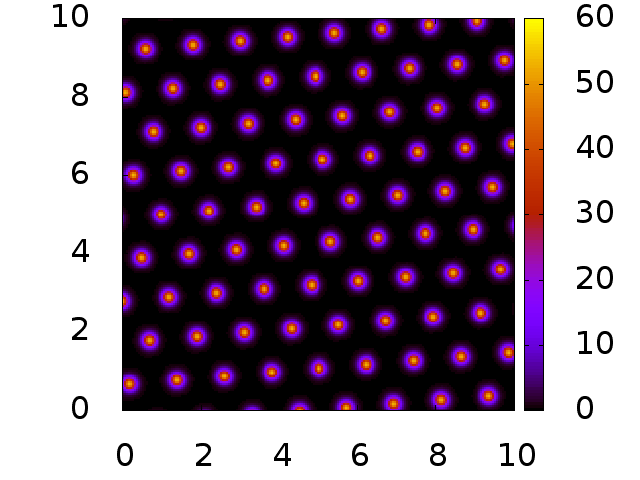} 
 \includegraphics[width=0.49\columnwidth]{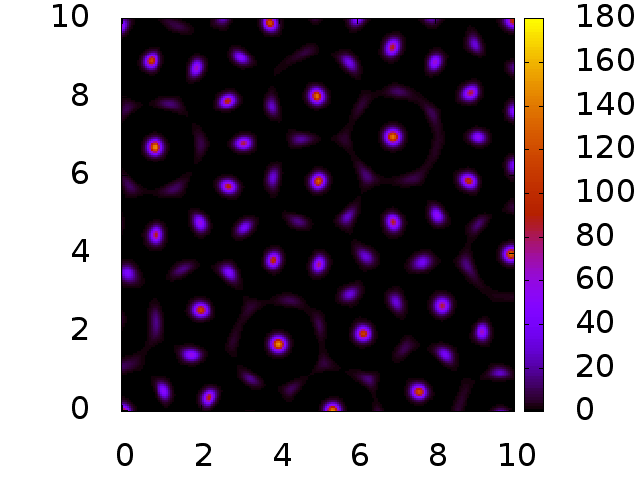} 
 \includegraphics[width=0.49\columnwidth]{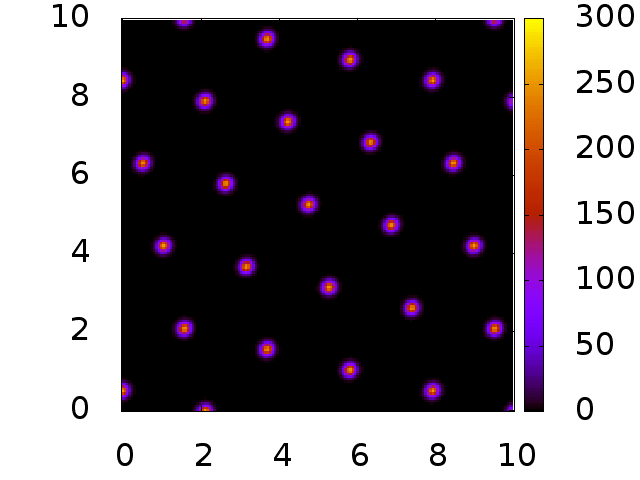} 
 \includegraphics[width=0.49\columnwidth]{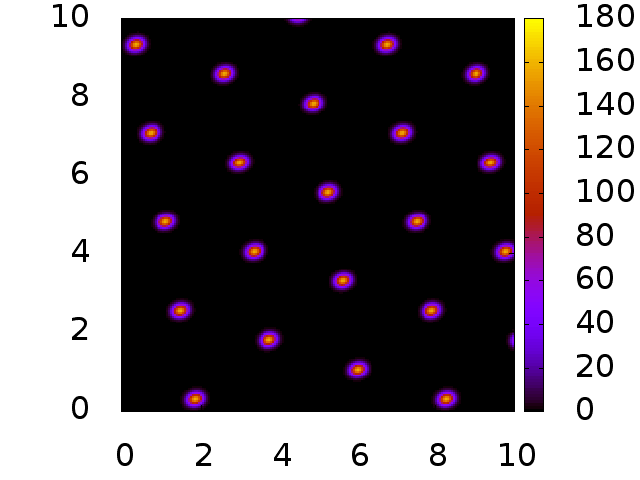} 
 
 \includegraphics[width=0.49\columnwidth]{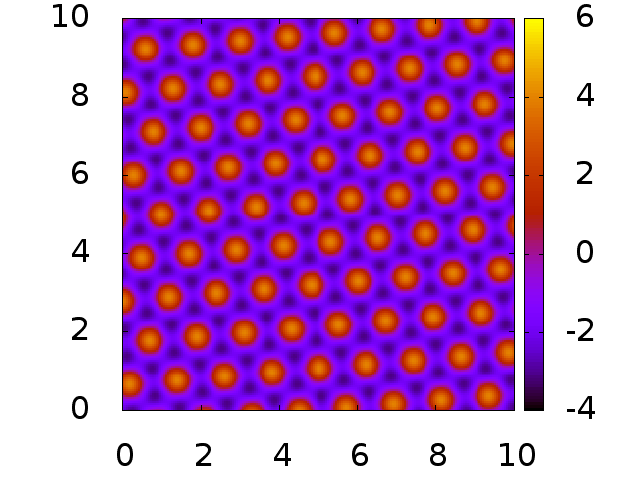} 
 \includegraphics[width=0.49\columnwidth]{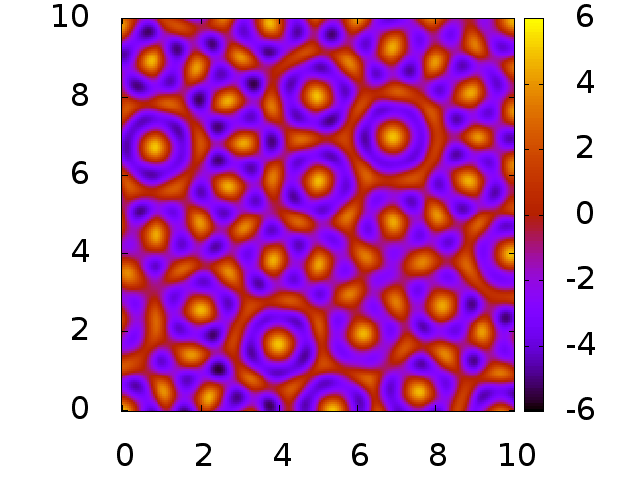} 
 \includegraphics[width=0.49\columnwidth]{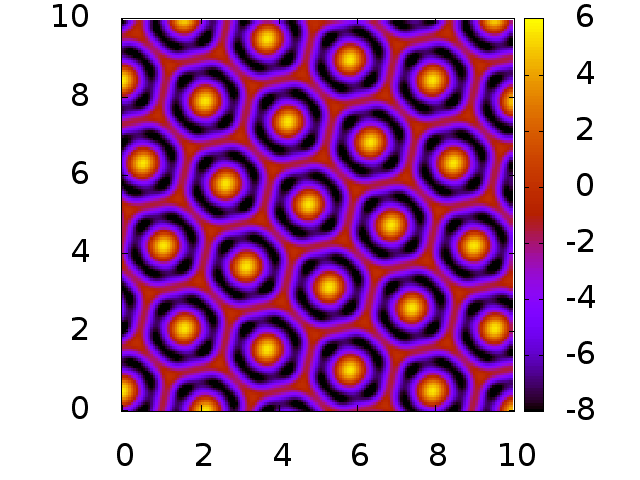} 
 \includegraphics[width=0.49\columnwidth]{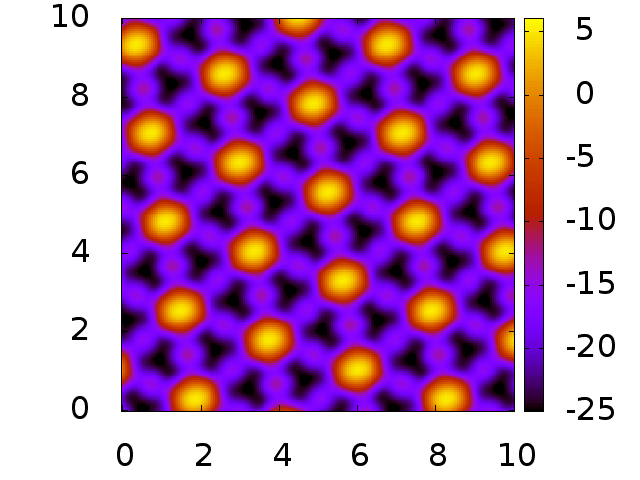} 
 
 \caption{\label{fig:profiles} (color online) Density profiles from DFT showing 
$R^2 \rho(\rr)$ (upper panels) and $\ln [R^2\rho(\rr)]$ (lower panels). Profiles 
for: (a) $(\rho_0R^2,A)=(3.6,0.5)$ (typical of the small length scale 
crystal~B), (b) $(3.5,0.76)$, (c) $(4.0,0.8)$ (both near the transition from 
crystal~A to crystal~B) and (d) $(2.7,2)$ (typical of the large length scale 
crystal~A). These state points are marked $\blacksquare$ in Fig.~\ref{fig:all}(a). 
The profiles in (b) show quasicrystalline ordering 
with numerous defects, while (c) reveals a network of connected 
density, indicating that the particles in this part of the crystal are fluid, 
and able to move throughout the system. There are also similar connected 
fragments in the disordered (b) profile, but because of the disorder, 
these do not percolate the system. 
 \vspace{-2ex}} 
 \end{figure*} 
 
Fig.~\ref{fig:all}(a) shows the equilibrium phase diagram calculated 
using Picard iteration \cite{Roth10} of the DFT Euler--Lagrange equation, starting either 
from the profile for a nearby state point or a 
uniform density profile with a small random value added to each point. 
As the fluid density is increased, the system freezes to form one of two 
distinct solid phases (Fig.~\ref{fig:profiles}): 
for larger values of $A$ the system forms crystal~A, a hexagonal crystal with a large 
lattice spacing, but for smaller values of $A$ 
it forms crystal~B, a hexagonal crystal with a much smaller lattice spacing. 
The red regions in Fig.~\ref{fig:all}(a) denote thermodynamic coexistence between two 
different phases at the same temperature, pressure and chemical potential.
 
To understand the phase diagram we study the structure and stability of a 
uniform liquid with density $\rho_0$ and $\Phi(\rr)\equiv 0$. We follow 
\cite{Evans79,Evans:TDGammaMolecP1979,ArEv04,ARTK12} and expand 
Eq.~\eqref{eq:DDFT} in powers of $\tilde{\rho}(\rr,t)\equiv\rho(\rr,t)-\rho_0$. 
Retaining only linear terms, we find that the growth/decay of different 
Fourier modes of wave number~$k$ follows $\hat{\rho}(k,t)=\hat{\rho}(k,0) 
\exp [\omega(k) t]$, where $\omega(k)$ satisfies the dispersion relation~\cite{ArEv04,ARTK12} 
 \begin{equation} 
 \omega(k)=-\Gamma k_BT \, k^2 (1 - \rho_0 \hat{c}(k)). 
 \label{eq:disp_rel} 
 \end{equation} 
Here $\hat{c}(k)$ is the Fourier transform of the pair direct correlation function; within RPA $\hat{c}(k)=-\beta \hat{V}(k)$, where $\hat{V}(k)$ is the Fourier transform of the pair potential in Eq.\ \eqref{eq:pair_pot}. In an equilibrium fluid the static structure factor $S(k)\equiv [1-\rho_0 \hat{c}(k)]^{-1}>0$ for all $k$; such a fluid is therefore stable \cite{HM}. Within RPA the two length scales in the pair potential lead, for certain ranges of parameter values, to a static structure factor $S(k)$ with two peaks. Fig.~\ref{fig:all}(b)--(d) shows that as $A$ increases the smaller $k$ peak in $S(k)$ grows and comes to dominate the larger $k$ peak. Fig.~\ref{fig:all}(e)--(g) shows analogous 
behavior of $\omega(k)$ at several points in or on the boundary 
of the linearly unstable region $\omega(k_{max})=0$, where $k_{max}$ is the wave 
number of the {\em higher} peak (blue dashed line in Fig.~\ref{fig:all}(a)): 
as $A$ increases the instability shifts from large $k$ 
(Fig.~\ref{fig:all}(g)) to small $k$ (Fig.~\ref{fig:all}(e)). The short 
and long length scales are simultaneously marginally stable at $A=1.067$ and 
$\rho_0R^2=2.95$ (Fig.~\ref{fig:all}(f)); this point lies on the pink dotted 
line in Fig.~\ref{fig:all}(a) corresponding to a pair of equal height peaks 
in $\omega(k)$. Above (below) this line, the peak at smaller 
(larger) wave number $k$ is higher, indicating that the longer (shorter) 
length scale density fluctuations grow the fastest. The black double dotted 
lines indicate the location of $\omega(k_{max})=0$ for the {\em lower} peak in 
$\omega(k)$. When the system is quenched from a stable liquid 
state to a state point with density $\rho_0$ above the blue dashed line, 
certain wave numbers will grow as described by $\omega(k)$. 
 
Fig.~\ref{fig:profiles}(c), shows the density 
profile of the larger lattice spacing crystal~A phase for a state point 
not far from the transition to the smaller lattice spacing crystal~B phase. 
However, the panel below displaying $\ln [R^2\rho(\rr)]$ reveals an interconnected 
network of channels between the density peaks. The particles contributing to 
this part of the density profile are fluid in the sense that they can move 
freely throughout the whole system, unlike the majority of the particles that 
are located in density peaks at multiply occupied lattice sites. {This 
is the crystal-liquid (CL) state.} This state minimizes the  
free energy for $A>A_\mathrm{co}$, where $A_\mathrm{co}$ is the value at 
coexistence. {Interfaces between the different phases in 
Fig.~\ref{fig:profiles} are present whenever these coexist 
(cf.\ \cite{Oettel12, MNT90}), but these will be discussed elsewhere.} 

\begin{figure} 
 \includegraphics[width=0.49\columnwidth]{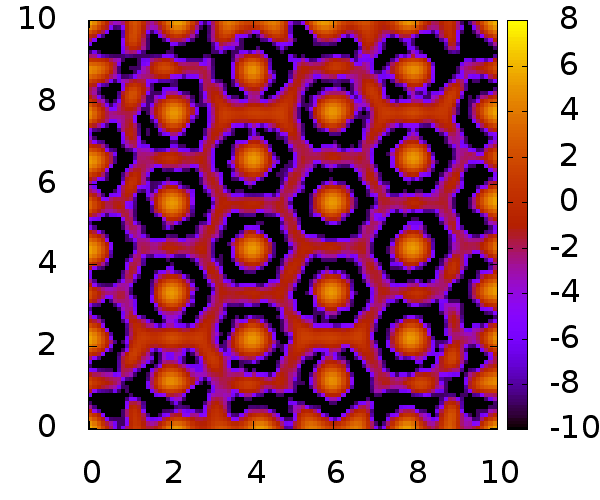} 
 \includegraphics[width=0.49\columnwidth]{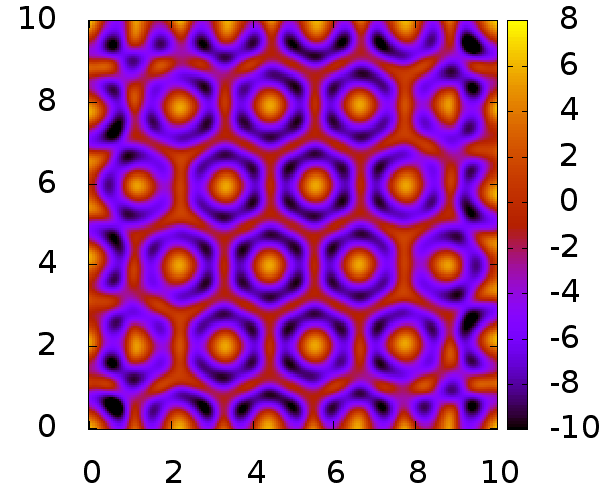} 
 \caption{\label{fig:DFTvBD} $\ln [R^2\rho(\rr)]$ for a system of $N=600$ particles 
with $(\rho_0R^2,A)=(4.0,0.8)$ confined in a square region of side $L=10 R$ obtained
from BD simulatios (left panel) and DFT (right panel). The 
system forms crystal~A with a density profile consisting of an array of 
peaks surrounded by a connected network within which the particles are free 
to move -- the CL state. 
} 
 \end{figure}

\begin{figure}[t]
 \includegraphics[width=0.9\columnwidth]{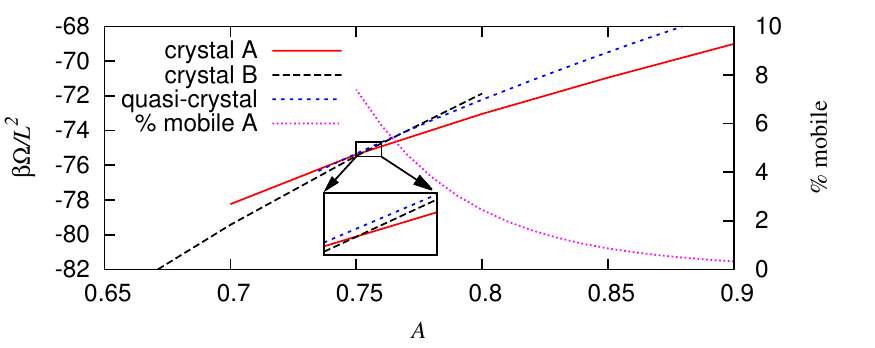} 
 \caption{Grand potential density for $\beta\mu=39$ as a 
function of $A$ for the two different crystal structures and 
the QC solution displayed in Fig.~\ref{fig:B}. There 
is a point where all three have almost the same free energy, but the 
QC solution is never the global minimum of the free 
energy (see inset). The crystal~A phase is of CL type throughout 
the range of $A$ shown. {We also display the \% of mobile
particles in the crystal A phase.}} 
 \label{fig:free_energy} 
 \end{figure}

To confirm the existence of the CL state we calculate the 
density profile for a system within a square confining potential 
$\Phi$ of size $L\times L$ with hard walls, and compare the results with 
Brownian dynamics (BD) simulations, i.e., simulations of $N$ particles 
evolving according to Eq.~\eqref{eq:EOM}. Averaging over the positions of 
the particles to calculate the density profile, we find remarkably good 
agreement between the DFT and the BD results (Fig.~\ref{fig:DFTvBD}). 
The resulting system thus consists of two dynamically distinct populations, in 
contrast to related systems \cite{Glaser_etal,ImRe04,ImRe06,Archer08} in which the
dynamics of all the particles are identical. {In Fig.~\ref{fig:free_energy} we
display, for $\beta\mu=39$, the percentage of mobile particles in crystal A as a 
function of $A$, obtained by integration over all portions of the density profile 
that are a distance $0.65R$ away from the centre of the density peaks. Particles 
contributing to this portion of the density are defined to be mobile. For 
$\beta\mu=0.39$, the two crystal phases coexist at $A_{\rm co}\approx0.75$; the 
proportion of mobile particles increases rapidly as $A\to A_{\rm co}^+$ and reaches 
over 7\% at this value of the chemical potential. In fact, as $A$ is further 
decreased it is this growing number that triggers the formation of the 
smaller length scale crystal: these mobile particles freeze to form the 
extra peaks of crystal~B. }
 
\begin{figure}
 \includegraphics[width=0.53\columnwidth]{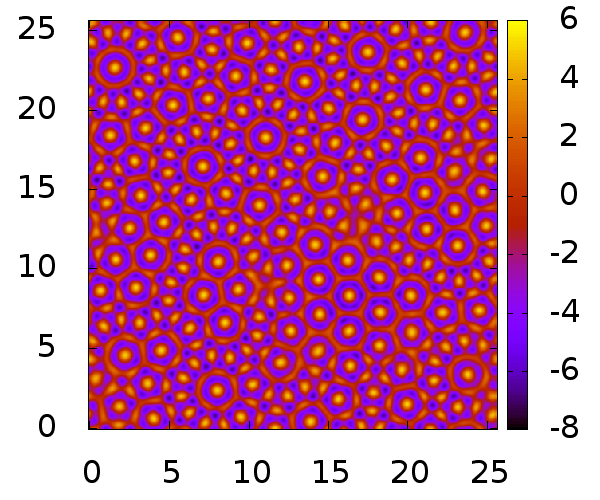} 
 \includegraphics[width=0.45\columnwidth]{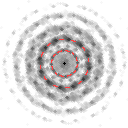} 
 
 \includegraphics[width=0.53\columnwidth]{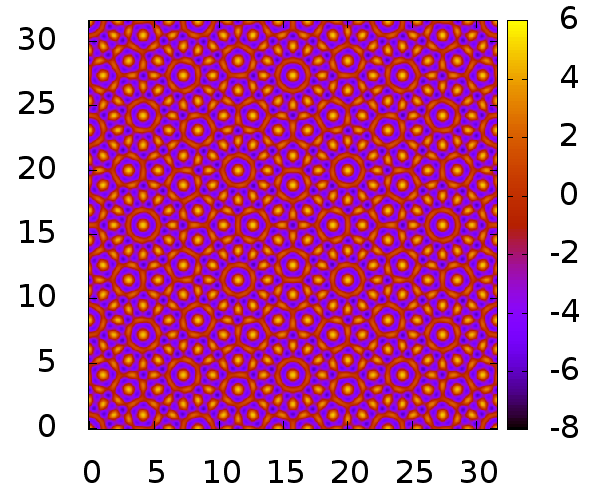} 
 \includegraphics[width=0.45\columnwidth]{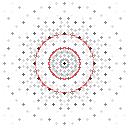} 
 \caption{Left: $\ln [R^2\rho(\rr)]$ from DFT, for $(\rho_0R^2,A)=(3.5,0.8)$. 
Right: the corresponding Fourier transforms. The 12-fold symmetry is indicative 
of QC ordering. The upper 
profile was obtained from 
random initial conditions, while the lower one was started from initial 
conditions with QC symmetry.} 
 \label{fig:B} 
  \end{figure}

\emph{Observation of metastable QCs:} A striking aspect of the phase 
diagram in Fig.~\ref{fig:all}(a) is that the phase transition 
between the two different crystal phases (thin red region) is well 
away from where the two peaks in the dispersion relation have the same 
height (pink dotted line). A uniform system quenched to the region above the 
coexistence of the two crystal phases but below this line 
will initially generate small length scale density 
fluctuations and the system behaves as if it were going to form crystal~B. 
However, the true minimum of the free energy is the larger length scale 
crystal. Thus, as growing density fluctuations reach the nonlinear regime, the 
system seeks to go to the longer length scale structure but the smaller length 
scale imprinted from the linear growth regime leads to frustration. Sometimes 
the system is able to evolve to the larger length scale crystal; at other times 
it stays stuck in the metastable small length scale crystal~B structure. 
However, often the system forms a state with density peaks on both length 
scales, but no long range order. In Fig.~\ref{fig:B} we display two rather 
striking density profiles calculated near state point 2b in Fig.~\ref{fig:all}(a).  
The upper profile was calculated using Picard iteration 
starting from random initial conditions. The density profile has many defects, 
but it has definite quasicrystalline ordering, as can be seen from the 
corresponding Fourier transform. The lower panels in Fig.~\ref{fig:B} show 
a defect-free QC approximant, started from carefully chosen initial 
conditions. The two wavenumbers $k_1R=3.2$ and $k_2R=6.0$ corresponding 
to the maxima in $\omega(k)$ are indicated in the Fourier transforms. 
 
The Picard iteration of the Euler--Lagrange equation corresponds to fictitious 
dynamics since it does not conserve the total number of particles in the 
system, $N\equiv\int\dr\rho(\rr)$. The true dynamics is governed by the DDFT 
Eq.~\eqref{eq:DDFT}. Evolving this equation is much slower, but in most 
cases the same qualitative behavior is observed. The supplementary material
below shows time-dependent QC formation obtained using DDFT. The conserved DDFT
dynamics does however lead to a higher likelihood of getting stuck in the  
crystal B structure formed in the initial linear growth regime. For 
$\beta\epsilon=1$, $R_s/R=1.855$ the QCs we find are never the minimum free  
energy state (Fig.~\ref{fig:free_energy}). The QC state in Fig.~\ref{fig:B}  
remains stable against small perturbations
for $1.77 <R_s/R< 2.18$, but we have not calculated the full  
phase diagram for $R_s/R\neq1.855$ (at $R_s/R=1.885$ the two marginally stable  
wave numbers (Fig.~\ref{fig:all}(f)) are very close to the ratio  
$2\cos(\pi/12)=1.93$, favoring 12-fold QCs). We believe it may be  
possible to use nonlinear dynamics techniques~\cite{Rucklidge2012} to compute  
the stability properties of these states by reducing the DDFT description in 
Eq.~\eqref{eq:DDFT} to a phase field crystal model, 
cf.~\cite{Elder2002,Huang2010a,Teeffelen2009,Emmerich2012,ARTK12}. We expect 
that the observed QC formation mechanism (linear growth of one length 
scale, but nonlinear selection favoring another) may well apply more generally. 
 
Acknowledgement: This work was supported in part by the National Science 
Foundation under grant DMS-1211953 (EK). We are grateful for discussions with 
R. Lifshitz and P. Olmsted.

\newpage

 \section{Supplementary material}

\begin{figure*}[h]

\noindent
\includegraphics[width=2.3in]{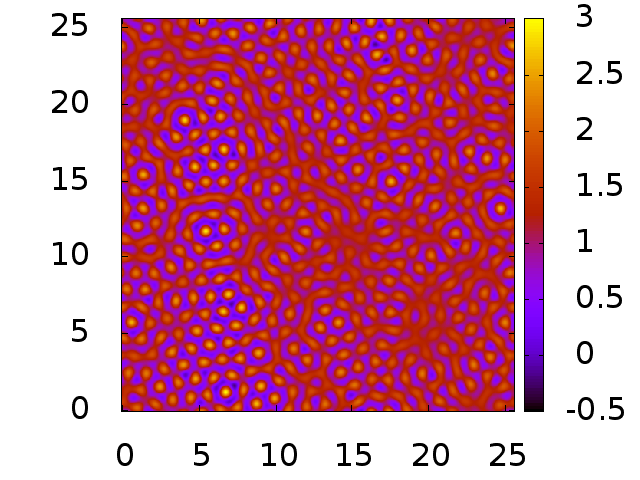} 
\includegraphics[width=2.3in]{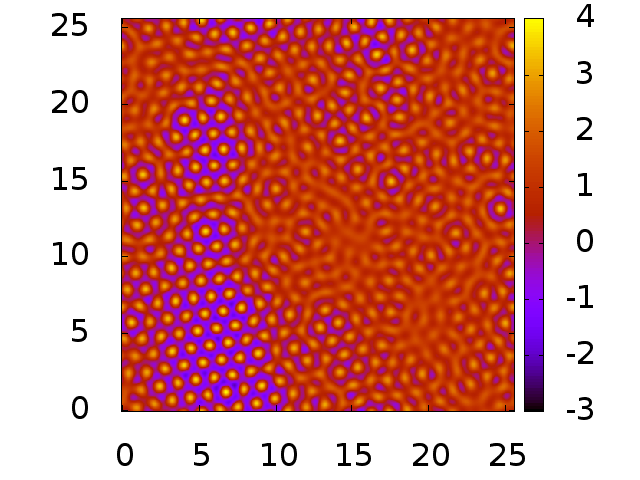}
\includegraphics[width=2.3in]{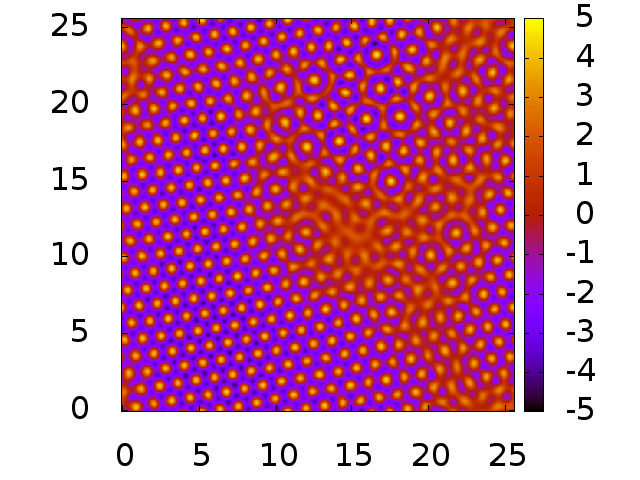} 

\noindent
\includegraphics[width=2.3in]{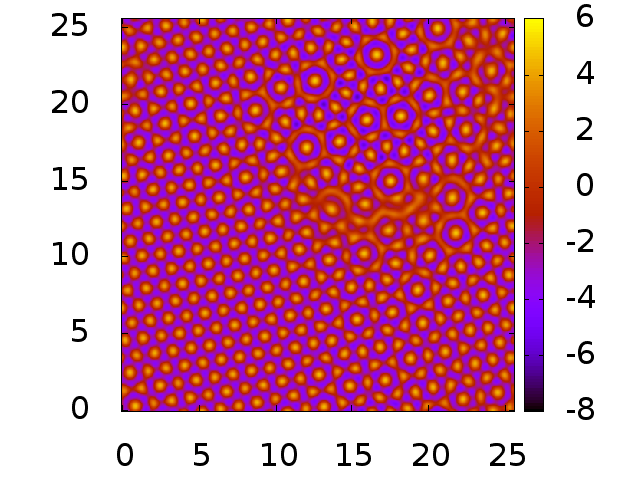} 
\includegraphics[width=2.3in]{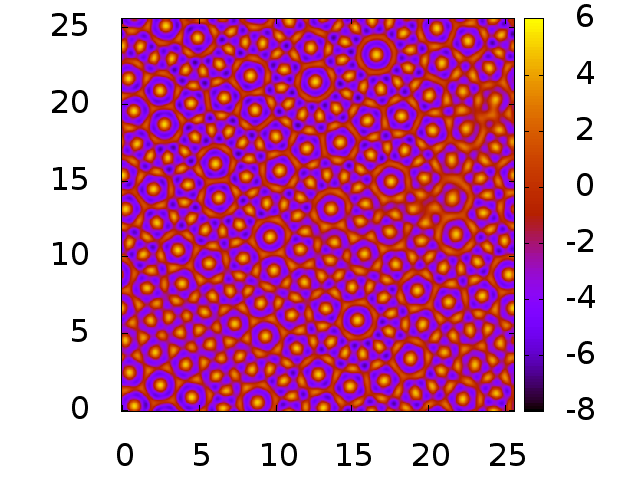} 
\includegraphics[width=2.3in]{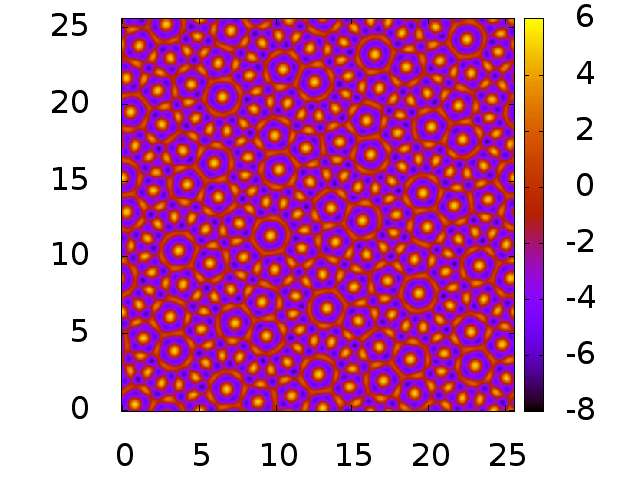} 

\caption{Time series of profiles $\ln[R^2\rho(\rr)]$ obtained via Picard iteration, for $A=0.8$ and $\rho R^2=3.5$. These show the evolution towards the equilibrium for the same state point as the results displayed in the upper panel of Fig.~4 of our Letter. The panels along the top row, from left to right, correspond to the times $t=30$, 32 and 35; and along the bottom row to $t=40$, 50, 200. Note that the system first forms (at time $t\approx30$) the small length scale crystal. It then tries to form the longer length scale crystal. However, due to the fact that it already has the small length scale imprinted on the system, it cannot form a perfect long length scale crystal and ends up forming a disordered system with regions of QC ordering -- see the final profile at time $t=200$, or the profile in Fig.~4 of the accompanying manuscript.}
   \label{fig:Picard}
\end{figure*}

\begin{figure*}

\noindent
\includegraphics[width=2.3in]{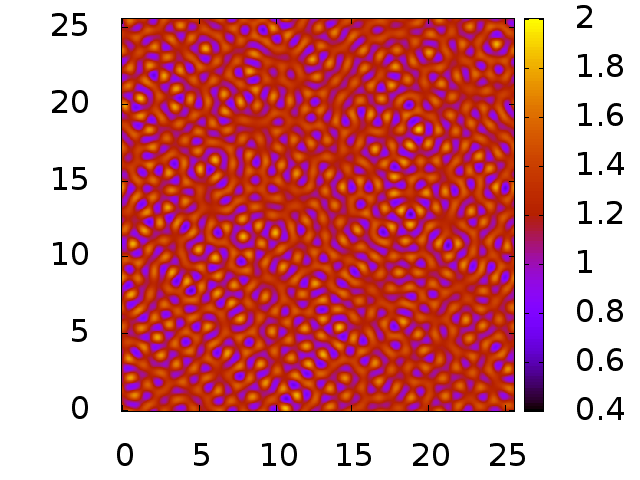} 
\includegraphics[width=2.3in]{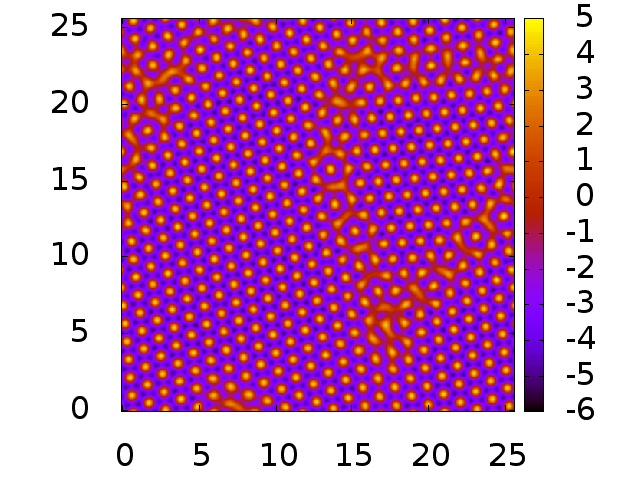} 
\includegraphics[width=2.3in]{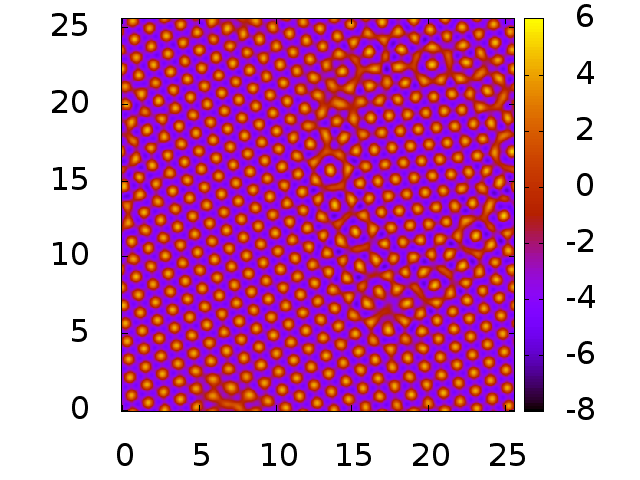} 

\noindent
\includegraphics[width=2.3in]{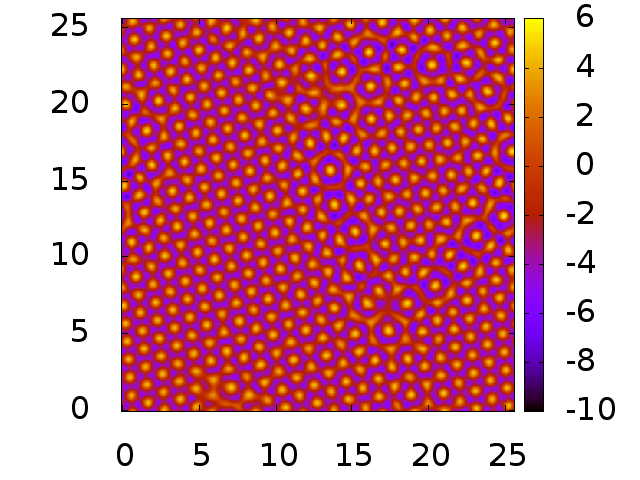} 
\includegraphics[width=2.3in]{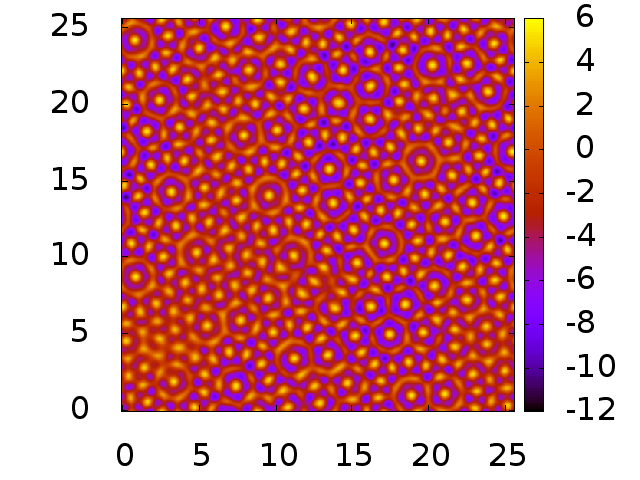} 
\includegraphics[width=2.3in]{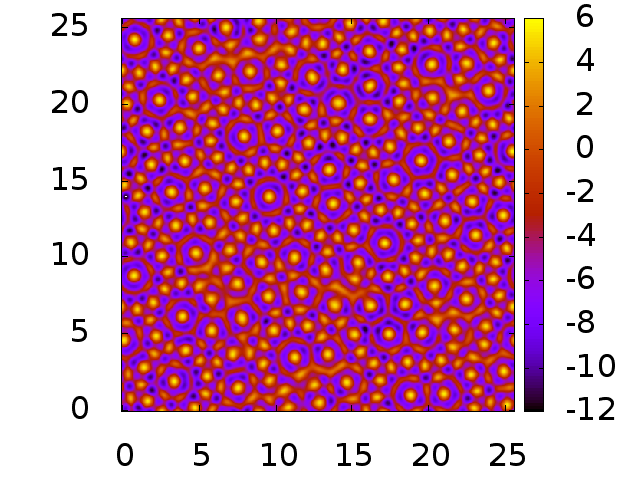} 

\caption{Time series of profiles $\ln[R^2\rho(\rr)]$ obtained from DDFT, for $A=1.067$ and $\rho R^2=3.5$. The panels along the top row, from left to right, correspond to the times $t/\tau_B=t^*=1$, 2 and 5; along the bottom row the panels correspond to $t^*=10$, 20 and 40, where $\tau_B=\beta R^2/\Gamma$ is the Brownian timescale.} 
   \label{fig:DDFT}
\end{figure*}

\begin{figure*}
   \centering
   \includegraphics[width=3in]{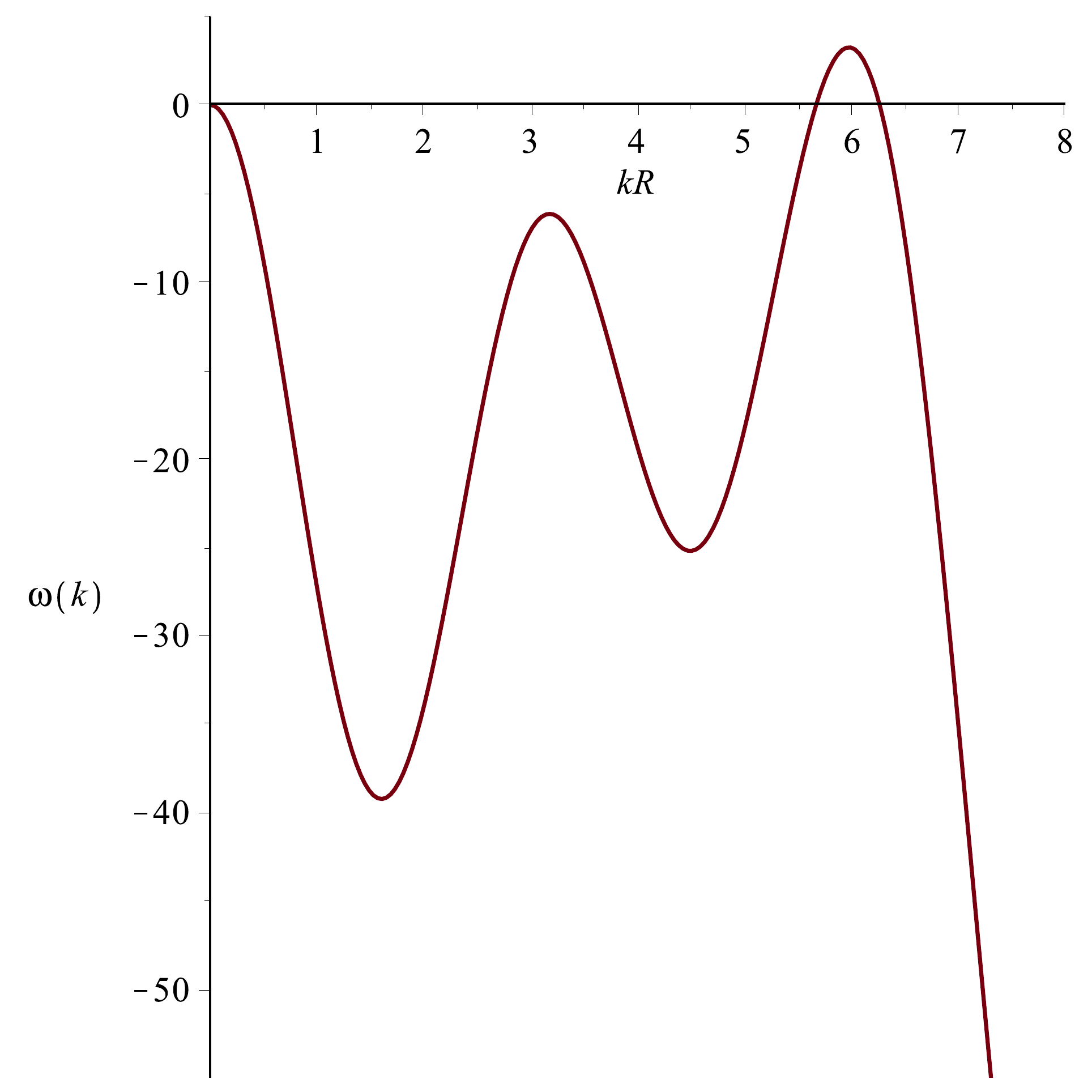} 
    \includegraphics[width=3in]{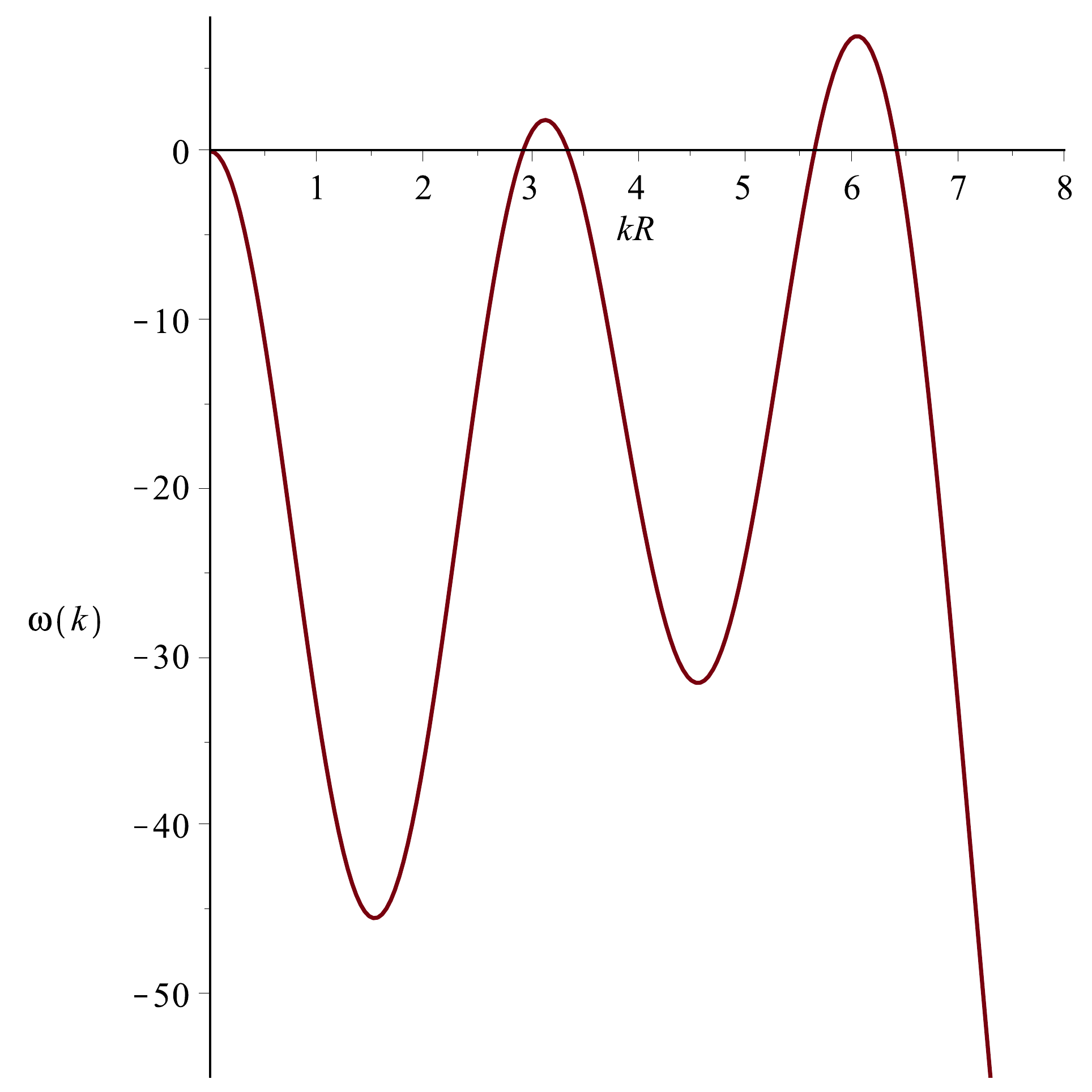} 
   \caption{The dispersion relation at the state point $A=0.8$ and $\rho R^2=3.5$ (left hand plot), where the QC state displayed in Fig.~4 of our Letter is calculated. From this we see that in the linear regime, only one mode (the smaller length scale) is unstable. Density profiles from the Picard iteration to equilibrium for this state point are displayed in Fig.~\ref{fig:Picard} above. Right hand plot: the dispersion relation for the state point $A=1.067$ and $\rho R^2=3.5$, which corresponds to the results in Fig.~\ref{fig:DDFT} above.}
   \label{fig:disp_rel1}
\end{figure*}

In this supplementary material, we display examples the dynamics of quasicrystal (QC) formation, obtained (i) via Picard iteration of the DFT (Fig.~\ref{fig:Picard}) and (ii) from dynamical density functional theory (DDFT) (Fig.~\ref{fig:DDFT}). In Fig.~\ref{fig:Picard} we display snapshots of the logarithm of the density profile $\ln[R^2\rho(\rr)]$ as the system evolves in time, for the state point $A=0.8$ and $\rho R^2=3.5$. The initial time $t=0$ profile corresponds to a uniform density plus a small amplitude random value everywhere. The early time linear growth regime produces {\bf one} length scale, as can be observed from Fig.~\ref{fig:Picard} and also from the dispersion relation, displayed in Fig.~\ref{fig:disp_rel1}. This leads to the formation of the small length-scale crystal B phase -- see, e.g.\ the middle and right hand panels in the top row of Fig.~\ref{fig:Picard}. However, over time, starting from the grain boundary and regions where defects are present, the longer length scale in the system appears, leading to the formation of the QC. This occurs when the dynamics of the system is well away from the linear regime.

In Fig.~\ref{fig:DDFT} we display snapshots of $\ln[R^2\rho(\rr)]$ as the system evolves in time, for the state point $A=1.067$ and $\rho R^2=3.5$ (note that the linear instability line is at $\rho R^2=2.95$, thus this state point is quite a deep quench). The dynamics we obtain from the DDFT displayed in Fig.~\ref{fig:DDFT} is very similar to that obtained from the Picard iteration in Fig.~\ref{fig:Picard}. The DDFT dynamics is for over-damped Brownian particles, as described in our Letter (and references therein). The Picard iteration used to generate Fig.~\ref{fig:Picard} does not conserve particle number. It is, however, much faster than the full DDFT and gives qualitatively similar results -- compare Figs.~\ref{fig:Picard} and \ref{fig:DDFT}. In Fig.~\ref{fig:disp_rel1} we also display the dispersion relation for the state point $A=1.067$ and $\rho R^2=3.5$, which corresponds to the results in Fig.~\ref{fig:DDFT}. Here we see that the system is linearly unstable at two distinct wavelengths. However, the peak corresponding to the smaller length scale (large $k$) is much higher than that of the longer length scale (small $k$) and so during the early time linear growth regime after the system is quenched, the smaller length scale grows much faster, so that there is no sign of the longer length scale in the early time density profile displayed in the top left hand panel of Fig.~\ref{fig:DDFT}.

 \end{document}